\documentclass[10pt,citeautoscript,floatfix,aps,prb,twocolumn,superscriptaddress,longbibliography]{revtex4-2}

\usepackage[english]{babel}
\usepackage{amsmath}
\usepackage{amssymb}
\usepackage{graphicx}
\usepackage[colorlinks=true, allcolors=blue]{hyperref}
\usepackage{physics}

\begin{document}

\title{%
Local topological markers for Chern insulators in ribbon geometry
}

\author{Maks Rep\v{s}e}
\email{maks.repse@fmf.uni-lj.si}
\affiliation{Faculty of Mathematics and Physics, University of Ljubljana, Slovenia}

\author{Toma\v{z} Rejec}
\affiliation{Faculty of Mathematics and Physics, University of Ljubljana, Slovenia}
\affiliation{Jožef Stefan Institute, Jamova 39, 1000 Ljubljana, Slovenia}

\author{Jernej Mravlje}
\affiliation{Jožef Stefan Institute, Jamova 39, 1000 Ljubljana, Slovenia}
\affiliation{Faculty of Mathematics and Physics, University of Ljubljana, Slovenia}

\date{\today}

\begin{abstract}
Local topological markers are used to characterize Chern insulators in the presence of spatial inhomogeneities, such as boundaries and disorder. In this paper, we study the local Chern marker in systems with partial translational symmetry.  We express the local Chern marker in the hybrid position-momentum basis for both open and periodic boundary conditions. We calculate the local Chern marker for a Haldane model ribbon. We show that the behavior at the two boundaries is qualitatively different from fully open geometries. We further compare the local Chern marker with the local St\v{r}eda marker and show agreement in the bulk and small deviations at the boundaries that diminish with increasing system size. The correspondence between the two markers remains good if disorder is introduced, provided its magnitude remains below large values that cause substantial change of the Chern number due to Anderson physics.
Finally, by exploiting the numerical efficiency due to partial translational symmetry, we study equilibrium critical behavior and the Kibble-Zurek mechanism in a weakly disordered Qi–Wu–Zhang Chern insulator. We extract relevant scaling exponents from the local Chern marker configuration and show that they converge to the analytically predicted values with increasing system size.
\end{abstract}

\maketitle

\section{Introduction}\label{sec:intro}
Local topological markers have been proven to be a useful tool for characterizing inhomogeneous topological systems. They were originally introduced in Chern insulators and used to recover the bulk Chern number $C$ in systems without translational symmetry, such as  disordered~\cite{prodan_entanglement_2010} and finite~\cite{bianco_mapping_2011} systems. The local Chern marker~(LCM), introduced in Ref.~\cite{bianco_mapping_2011}, is, in a sense, a spatially resolved density of $C$. Since its introduction, LCM has been connected to several observable quantities, such as orbital magnetization~\cite{bianco_orbital_2013}, anomalous Hall conductivity~\cite{marrazzo_locality_2017,dornellas_quantized_2022} and circular dichroism~\cite{tran_probing_2017}.

The wide adoption of LCM and related quantities has primarily focused on systems that feature no translational symmetry. That is, systems within open boundary conditions in all directions~(fully-OBC) and/or disorder that breaks translational symmetry even in systems with periodic boundary conditions~(PBC). In contrast, systems with partial translational symmetry have received far less attention. Examples of such systems include ribbons~(2-dimensional systems that are infinite in one and finite in the other direction) and systems with inhomogeneities or disorder oriented in stripes. Moreover, partial translational symmetry is a natural description of interfaces~\cite{irsigler_interacting_2019} and domain walls between regions with different $C$, bulk systems in a homogeneous external magnetic field in the Landau gauge and periodically driven Floquet Chern insulators~\cite{privitera_quantum_2016}.

Here, we build on the work of Drigo and Resta~\cite{drigo_chern_2020} and use a formula for calculating LCM in systems with partial translational symmetry using the position-momentum~(hybrid) basis. In Sec.~\ref{sec:markers} we review the derivation of the formula and present an approach for its numerical evaluation within both periodic and open boundary conditions. In Sec.~\ref{sec:haldane-ribbon} we explore the behavior of LCM in ribbon geometry and compare it to a different topological marker, the local St\v{r}eda marker~\cite{markov_locality_2024}, motivated by the experimentally measurable response of the system to an external uniform magnetic field. We show data that  suggest that for moderately strongly disordered systems the two markers are equal in the thermodynamic limit. In Sec.~\ref{sec:KZM} we demonstrate the utility of partial translational symmetry
to study slow quenches over topological phase transitions. We accurately determine how correlation length scales with the quench duration in a model Chern insulator.
In Sec.~\ref{sec:conclusion} we present our conclusions.

Throughout the article, we set the reduced Planck constant $\hbar$ and the lattice constant $a$ to 1.

\section{Local topological markers in position-momentum basis}\label{sec:markers}
\subsection{Local Chern marker}
\label{subsec:markers-chern}
We assume spinless and non-interacting electrons and start with the expression for the LCM in the position basis~\cite{bianco_mapping_2011,prodan_entanglement_2010}
\begin{equation}\label{eq:LCM-rr}
    c(\vb r)=2\pi i\sum_{\alpha}\expval{P\comm{-i\comm{\hat x}{P}}{-i\comm{\hat y}{P}}}{\vb r,\alpha},
\end{equation}
where~$\vb r=(x,y)$ is the lattice vector and~$\alpha$ denotes the orbital index. We denote the number of unit cells in $x(y)$ direction as~$N_{x(y)}$. The Fermi projector~$P=\sum_{E_{\psi}<\mu}\ket{\psi}\bra{\psi}$ projects onto the occupied subspace that is spanned by single-electron eigenstates~$\ket{\psi}$ of the system's Hamiltonian with energies~$E_{\psi}$ below the Fermi level~$\mu$. 

Assuming translational invariance in the~$y$ direction, the Fermi projector can be decomposed into~$N_y$ independent blocks, indexed by momenta~$k_y$ from the Brillouin zone
\begin{equation}\label{eq:projector-xky}
    P=\bigoplus_{k_y}P(k_y).
\end{equation}
The first commutator in Eq.~\eqref{eq:LCM-rr} decomposes into a direct sum of commutators~$-i\comm{\hat x}{P}\to\bigoplus_{k_y}-i\comm{\hat x}{P(k_y)}$ and the second commutator becomes $-i\comm{\hat y}{P}\to\bigoplus_{k_y}\partial_{k_y}P(k_y)$~\cite{prodan_disordered_2011}.
This yields the expression for the LCM in position-momentum basis~\cite{drigo_chern_2020}
\begin{equation}\label{eq:LCM-rk}
    \begin{split}
        c(x)=\frac{2\pi i}{N_y}\sum_{k_y}\bigg\{ \sum_{\alpha} \bra{x,\alpha} &P(k_y) \\ \big[-i\comm{\hat x}{P(k_y)}\,,\,&\partial_{k_y}P(k_y)\big] \ket{x,\alpha} \bigg\}.
    \end{split}
\end{equation}

Since $P(k_y)$ is periodic in $k_y$, its derivative can be calculated numerically using the spectral derivative~\cite{sirca_computational_2018}. We use the Fourier transform $\mathcal F$ to determine the series of Fourier components ${\cal P}_n$ of the projector. The $k_y$ derivative of $P(k_y)$ can then be calculated as the inverse Fourier transform of the series
\begin{equation}
    \partial_{k_y}P(k_y)\to\mathcal{F}^{-1}\{in{\cal P}_n\big\}
\end{equation}
If $N_y$ is even, the series runs from $n=-N_y/2$ to $N_y/2-1$~\footnote{In this case, the component $n=-N_y/2$ should be set to 0. See Ref.~\cite{sirca_computational_2018} for details}. If $N_y$ is odd, the series runs between $-(N_y-1)/2$ and $(N_y-1)/2$.

The calculation of the commutator $\comm{\hat x}{P(k_y)}$ depends on the boundary condition in the $x$ direction. In order to build some intuition it is useful to first consider a system with OBC in the $x$ direction, in which the position operator $\hat x$ is well defined. The $k_y$ component of the projector~\eqref{eq:projector-xky} can be written as a matrix in the position basis
\begin{equation}\label{eq:proj-pos-basis}
    P(k_y)=\sum_{x,x'}\ket{x}P_{xx'}(k_y)\bra{x'},
\end{equation}
where $\ket{x}$ is an eigenstate of the position operator $\hat x\ket{x}=x\ket{x}$ and $P_{xx'}(k_y)=\matrixel{x}{P(k_y)}{x'}$. The commutator $\comm{\hat x}{P(k_y)}$ acting on a position eigenstate $\ket{x_0}$ gives
\begin{equation}\label{eq:xP-infinite}
    \comm{\hat x}{P(k_y)}\ket{x_0}=\sum_x(x-x_0)P_{xx_0}(k_y)\ket{x}.
\end{equation}

Within PBC, the position operator is not well defined. However, Eq.~\eqref{eq:xP-infinite} can still be used if we modify the distances $x-x_0$ to account for periodicity. We define a symmetric $N_x\times N_x$ Toeplitz matrix of distances between lattice sites $\Delta_x$ with the first row
\begin{equation}\label{eq:toeplitz}
    \begin{pmatrix}
        0 & 1 & 2 & \cdots & M & -M & \cdots & -2 & -1
    \end{pmatrix}
\end{equation}
where $M=(N_x-1)/2$ and $N_x/2$ for $N_x$ odd and even, respectively. In the latter case, $M$ needs to be removed from the matrix and $-M$ set to 0. Eq.~\eqref{eq:xP-infinite} reduces to an element-wise product of matrices, denoted by $\odot$,
\begin{equation}\label{eq:comm-pbc}
    -i\comm{\hat x}{P(k_y)}\to-i\Delta_x\odot P(k_y).
\end{equation}
An alternative derivation of Eqs.~\eqref{eq:LCM-rk} and~\eqref{eq:comm-pbc} is provided in Appendix~\ref{app:derivation}.

Eq.~\eqref{eq:comm-pbc} can also be used when calculating commutators $\comm{\hat x}{P}$ and $\comm{\hat y}{P}$ in a fully local basis, assuming PBC in both directions. In that case, the formula produces results that are equivalent to those obtained by the formula presented in Ref.~\cite{prodan_entanglement_2010}. Furthermore, it could be useful when evaluating commutators of the type $\comm{\hat x}{P}$, in order to calculate the local $\mathbb Z_2$ topological marker in two-dimensional systems~\cite{bau_local_2024} or local topological markers in odd spatial dimensions~\cite{hannukainen_local_2022} within PBC.

\subsection{Local St\v{r}eda marker}
While LCM is an excellent tool for diagnosing topological properties in theoretical and numerical applications, its experimental implementation remains unclear. Another local topological marker for Chern insulators that could readily be measured in an experiment, is the local St\v{r}eda marker
\begin{equation}\label{eq:local-streda-def}
    c_S(\vb r)=\phi_0\pdv{n(\vb r)}{\phi},
\end{equation}
where $n(\vb r)$ is the local electron density, $\phi$ an external uniform magnetic flux per unit cell, $\phi_0=h/e_0$ the magnetic flux quantum and $e_0$ and $h$ unit charge and Planck constant, respectively.~(Unless stated otherwise, all fluxes in the remainder of this paper are given per unit cell.) The derivative must be evaluated in the limit $\phi\to0$ and at a fixed chemical potential $\mu$. The local St\v{r}eda marker has been connected to both LCM and orbital magnetization~\cite{bianco_orbital_2013}. Recently,  non-equilibrium behavior of local Chern and St\v{r}eda markers in systems within fully-OBC has also been investigated~\cite{markov_locality_2024}. We calculate $c_S$ in the hybrid basis simply as
\begin{equation}\label{eq:local-streda-1d}
    c_S(x)=\phi_0\pdv{\big(\sum_{k_y}{P_{xx}(k_y)\big)}}{\phi}
\end{equation}

\section{Haldane ribbon}\label{sec:haldane-ribbon}
\begin{figure}[ht]
        \includegraphics[width=3.25 cm]{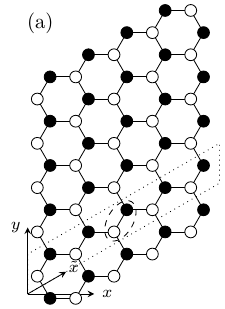}
    \hfill
        \includegraphics[width=5.2 cm]{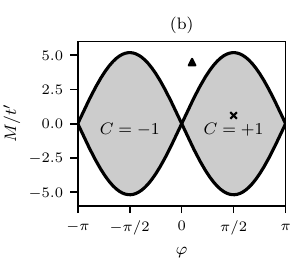}
    \caption{(a) The Haldane model ribbon: honeycomb lattice with white and black sites corresponding to on-site energies $\pm M$. The slice, marked with the dotted parallelogram, is the unit cell. The two `orbitals' are marked by the dashed ellipse. (b) Phase diagram of the Haldane model.}
    \label{fig:haldane-phase-diag}
\end{figure}

We first consider a Haldane model~\cite{haldane_model_1988} ribbon~(OBC in the $x$ and PBC in the $y$ direction) with zigzag boundaries. The model is a 2-dimensional tight-binding Hamiltonian on a honeycomb lattice~(Fig.~\ref{fig:haldane-phase-diag}(a)) that features staggered on-site energies $\pm M$, real nearest neighbor hopping with amplitude $t$ and complex next-nearest neighbor hopping $t'e^{i\varphi}$. We set the hopping amplitudes to $t=1$ and $t'=1/3$. We implemented the model using the Python library Kwant~\cite{groth_kwant_2014}.

The Haldane model features different insulating phases at half-filling. Depending on the parameters $M$ and $\varphi$, the insulator can either be trivial~($C=0$) or topological~($C=\pm1$) according to the phase diagram shown in Fig~\ref{fig:haldane-phase-diag}(b). Below, we consider the two points marked in the phase diagram. The triangle marks a point in the trivial phase with $M=1.5$ and $\varphi=\pi/10$ and the cross a point in the topological phase with $M=0.2$ and $\varphi=\pi/2$.

\begin{figure}[ht]
    \centering
    \includegraphics[width=\linewidth]{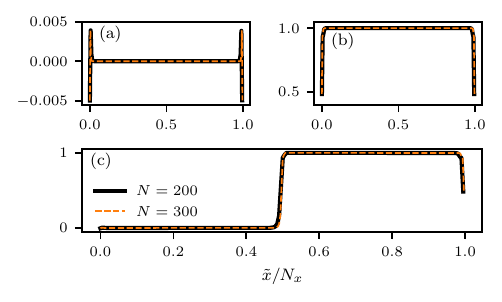}
    \caption{LCM configurations (a)~of the system in the trivial phase~(marked by the triangle in Fig.~\ref{fig:haldane-phase-diag}(b)), (b)~topological phase~(cross in Fig.~\ref{fig:haldane-phase-diag}(b)) and (c)~for a heterojunction, with the system's parameters corresponding to (a) in the left half and (b) in the right half. The two lines represent systems with different sizes $N=N_x=N_y$.}
    \label{fig:lcms}
\end{figure}

Configurations of the LCM, calculated using the procedure described in Sec.~\ref{subsec:markers-chern}, are shown in Fig.~\ref{fig:lcms}. Panel~(a) shows the configuration in the trivial phase~(point ``triangle'' in Fig.~\ref{fig:haldane-phase-diag}(b)). The marker is equal to 0 in the interior and features a small deviation at each boundary. The average of LCM over the entire system approaches $C$ with increasing system size.

In the topological phase~(point ``cross'', panel~(b)) LCM equals $C=+1$ in the bulk and deviates at the boundary. However, in contrast with the fully-OBC case~\cite{bianco_mapping_2011}, the boundary contribution does not compensate that of the bulk. Furthermore, the amplitude of the deviations at the boundaries is independent of system size, therefore the average of LCM over the entire system approaches $C$ as~$\sim1/N_x$, as was already demonstrated in Ref.~\cite{drigo_chern_2020}.

Panel~(c) shows LCM configuration for a ribbon heterojunction. The left half of the system corresponds to the point marked by the triangle and the right half to the point marked by the cross in Fig.~\ref{fig:haldane-phase-diag}(b). The marker is capable of clearly distinguishing between regions with different $C$.

We compute the local St\v{r}eda marker by including a homogeneous perpendicular magnetic field in the Hamiltonian via Peierls substitution. The hopping amplitude $t_{ij}$ between sites $i$ and $j$ at positions $\vb r_{i(j)}$ is multiplied by a phase factor
\begin{equation}
    t_{ij}\to t_{ij}\exp{-i\frac{2\pi}{\phi_0}\int_{\vb r_i}^{\vb r_j}\dd\vb r\vb A(\vb r)},
\end{equation}
where the line integral is along the shortest path between $\vb r_i$ and $\vb r_j$ and $\vb A=\big(0,\delta\phi/(\sqrt3/2)[x-x_0]\big)$ is the magnetic vector potential with $x_0$ chosen such that $x-x_0=0$ on the central line of the ribbon. The system is then diagonalized at zero and at a finite magnetic flux $\delta\phi$, keeping $\mu$ constant. Electron densities are calculated for both cases and subtracted in order to calculate $c_S$ via Eq.~\eqref{eq:local-streda-1d}. 

\begin{figure}[ht]
    \centering
    \includegraphics[width=\linewidth]{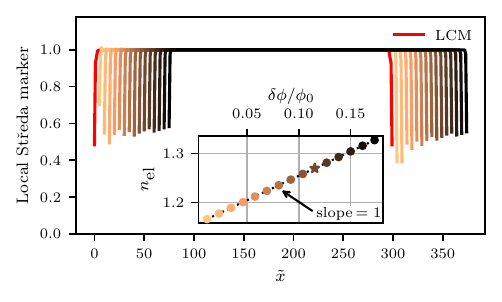}
    \caption{Local St\v{r}eda marker calculated at different $\delta\phi$~(shown in color scale). The $x$ coordinate is shifted at each flux to show boundary behavior. Red line shows the local Chern marker. The inset shows the average electron density $n_\textrm{el}$ in the ground state at each magnetic flux. Size of the system is $N_x=N_y=300$.}
    \label{fig:streda-B-convergence}
\end{figure}

Fig.~\ref{fig:streda-B-convergence} shows the behavior of the local St\v{r}eda marker at different $\delta\phi$ used in its calculation. The value of $c_S$ in the bulk is independent of $\delta\phi$ and equals the value of the bulk Chern number and LCM. This 
is in line with the behavior of the markers in the fully-OBC case, see Refs.~\cite{bianco_orbital_2013,markov_locality_2024}.

At the boundary, the value of $c_S$ converges with increasing $\delta\phi$. The $\delta\phi$ at which the value converges decreases with the system's size as $\delta\phi\sim1/N_y$~(see Movie in supplemental material~\cite{supplemental}). This means that the finite $\delta\phi$ needed for $c_S$ to converge at the boundaries is only a finite size effect. Once $c_S$ converges at the boundary, its value there is similar to LCM.

The inset in Fig.~\ref{fig:streda-B-convergence} shows the average electron density in the ground state at different $\delta\phi/\phi_0$. The dependence is linear with the slope equal to $C=1$. This is in accordance with the St\v{r}eda formula~\cite{streda_theory_1982}.

\begin{figure}[ht]
    \centering
    \includegraphics[width=\linewidth]{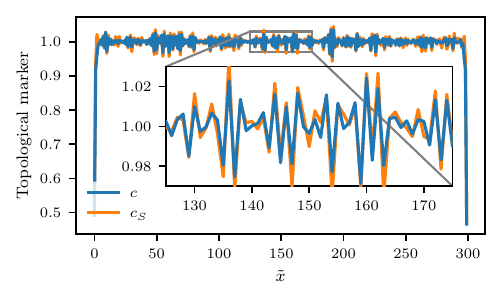}
    \caption{Configuration of LCM $c$~(blue line) and local St\v{r}eda marker $c_S$~(orange line) in the topological phase with on-site disorder. Disorder amplitude is $\delta=1$, size of the system is $N_x=N_y=300$ and $\delta\phi=30\,\phi_0/N_y$.}
    \label{fig:disorder-comp}
\end{figure}

We also studied the correspondence between the two markers when disorder is introduced to the system. At each $x$, a random number $\delta_x$ is sampled uniformly from the interval $[-\delta/2,\delta/2]$ and added to the on-site energy term. Importantly, $\delta_x$ does not vary along $y$ in order to preserve partial translational symmetry.  
The markers are calculated as before, with $\delta\phi$ set to $30\,\phi_0/N_y$~(marked by the star in the inset of Fig.~\ref{fig:streda-B-convergence}). In Fig.~\ref{fig:disorder-comp} we show results for disorder amplitude $\delta=1$. One sees the two markers agree closely. (Result with $\delta\phi=\phi_0/N_y$ is shown in Fig.~\ref{fig:disorder-comp-min-flux} in Appendix~\ref{app:disorder}.)  
The similarity breaks down for $\delta \gtrsim 3$, when the average of LCM in the bulk deviates substantially from $C=1$ due to Anderson physics~(see Figs.~\ref{fig:disorder-app-w3}--\ref{fig:disorder-anderson} in Appendix~\ref{app:disorder}).

\section{Kibble-Zurek mechanism in a disordered Chern insulator}\label{sec:KZM}

Local Chern marker has also been used to characterize non-equilibrium states after quantum quenches across topological phase transitions~\cite{ulcakar_kibble-zurek_2020,yuan_kibble-zurek_2024}. It was found that for slow quenches characterized by a duration $\tau$, the non-equilibrium state features a length scale that grows as the quench becomes slower, just as expected in systems with standard Landau order parameters. Small random disorder was introduced to reveal the length scale in real space. The length scale was proposed to be governed by the Kibble-Zurek mechanism, but
these predictions could be numerically tested for systems of moderate size only.
Here we take advantage of partial translational symmetry to show convergence of the results with the system size.

\subsection{Kibble-Zurek mechanism}
Kibble-Zurek mechanism~(KZM)~\cite{kibble_topology_1976,zurek_cosmological_1996} describes how the correlation length of a system~$\xi$ freezes-out during a slow quench across a critical point. It also predicts that $\xi$ at the end of the quench scales with the duration of the quench~(assuming a linear ramp)~$\tau$ as
\begin{equation}\label{eq:quench-xi-scaling}
    \xi\sim\tau^{1/(1+\nu z)},
\end{equation}
where~$\nu$ is the correlation length critical exponent and~$z$ the dynamical critical exponent~\cite{dziarmaga_dynamics_2010}.

As described below, the correlation length in a Chern insulator can be obtained from the LCM configuration. However, if the system is translationally invariant, the LCM configuration is homogeneous~(as is any other spatially resolved quantity). The correlation length can be revealed by introducing inhomogeneities like random disorder in the system. 
Since the disorder is needed only to reveal the underlying length scale (present in correlation functions or the Berry curvature~\cite{chen_scaling_2016}), it can be chosen in a way that preserves translational symmetry in one direction. This symmetry can then be used to speed up numerical calculations and allow for the study of larger systems.

\begin{figure*}[t]
    \centering
    \includegraphics{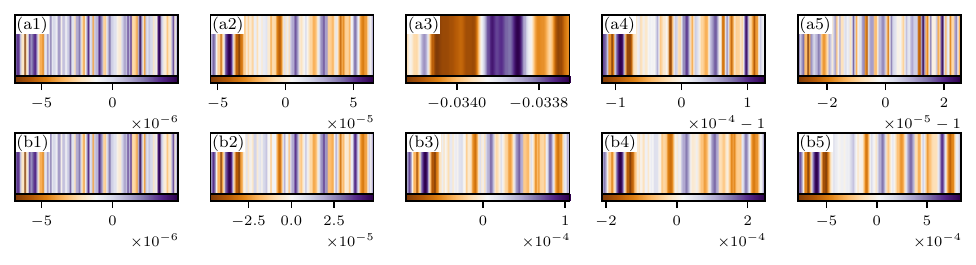}
    \caption{(a)~LCM in the ground state of an instantaneous Hamiltonian at~$u=-2.5$~(a1),~$-2.1$~(a2),~$-2.01$~(a3),~$-1.9$~(a4) and~$-1.5$~(a5). (b)~LCM during a linear quench from~$u=-2.5$ to~$u=-1.5$ with~$\tau=20$ at times~$t/\tau=0$~(b1),~$0.4$~(b2),~$0.49$~(b3),~$0.6$~(b4) and~$1$~(b5). The size of the system is~$N_x=N_y=300$, although only a third is shown for clarity.}
    \label{fig:profili}
\end{figure*}

\subsection{Model}
We use the Qi-Wu-Zhang model~\cite{qi_topological_2006} and assume PBC in both directions. The disorder~$\delta_{x}$ is invariant under translations in the~$y$ direction, making~$k_y$ a good quantum number. The Hamiltonian can therefore be written in block-diagonal form in the~$(x,k_y)$ basis as
\begin{equation}\label{eq:full-H}
    H_u=\bigoplus_{k_y}H_u(k_y),
\end{equation}
where~$k_y$ is an element of the Brillouin zone. Each block~$H_u(k_y)$ has the form~(the nearest neighbor hopping amplitude is set to 1)
\begin{equation}\label{eq:hk}
    \begin{split}
        H_u&(k_y)=\sum_{x=1}^{N_x} \left(\ket{x+1}\bra{x} \otimes\frac{\sigma_z+i\sigma_x}2 + {\rm h.c.}\right)+\\ &\sum_{x=1}^{N_x}\ket{x}\bra{x} \otimes[\cos k_y\sigma_z+\sin k_y\sigma_y+(u+\delta_{x})\sigma_z],
    \end{split}
\end{equation}
where~$x$ is the cell index and Pauli matrices~$\sigma_{x,y,z}$ act on a two-dimensional orbital space. The disorder~$\delta_{x}$ is uncorrelated and uniformly distributed on the interval~$[-\delta/2,\delta/2]$. We set disorder amplitude to $\delta=5\times10^{-4}$ throughout.

The phase diagram of the~(clean) QWZ model features the following insulating phases: The system is a trivial insulator with $C=0$ for $\abs{u}>2$ and a topological insulator with $C=-1$ and $+1$ for $-2<u<0$ and $0<u<2$, respectively.

\subsection{Critical behavior in the ground state}
We probe the critical ground state behavior of the system by diagonalizing the Hamiltonian~\eqref{eq:full-H}, calculating the Fermi projector and evaluating LCM via Eq.~\eqref{eq:LCM-rk}. The correlation length is determined as the distance over which the LCM autocorrelation function falls to 0~\cite{ulcakar_kibble-zurek_2020}. We focus on the behavior near the critical point at~$u_c=-2$.

Fig~\ref{fig:profili}(a) shows LCM profiles at different distances from the critical point. Panels~(a1) and~(a5) show LCM profiles far away from the critical point in the trivial and topological phases, respectively. The profiles feature small inhomogeneities -- regions where LCM deviates above~(blue) or below~(brown) its average. Correspondingly, the correlation length $\xi$ is also small. On approaching the critical point~[panels~(a2) and~(a4)], the inhomogeneities grow with largest features seen in the data closest to the critical point~[panel~(a3)]. Note that the Chern number, calculated as the average of LCM, is not precisely equal to $C=0$. This is due to a finite-size effect. In panels~(a4) and~(a5) the contribution directly proportional to disorder is filtered out~\cite{ulcakar_kibble-zurek_2020} using a Gaussian filter with the width 1~(lattice spacing).

\begin{figure}[ht]
    \centering
    \includegraphics[width=\linewidth]{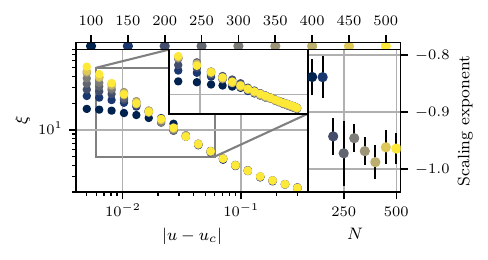}
    \caption{Left: Correlation length in the ground state at different distances from the critical point. Inset shows points in the linear regime~(note the log-log scale) used for fitting. Color scale represents system size~$N_x=N_y$. Right: Scaling exponent~$-\nu$ obtained by fitting a power law to the points show in the inset. Fit and standard deviation are obtained by bootstrap resampling.}
    \label{fig:groundstate}
\end{figure}

Fig.~\ref{fig:groundstate} shows the dependence of correlation length~$\xi$ in the ground state on the distance from the critical point~$\abs{u-u_c}$ on the trivial side. The result is averaged over 100 disorder realizations. The data clearly shows a power law behavior, with the maximal length reached limited by the system size. The inset shows points in the linear regime that are used to determine the scaling exponent~$-\nu$. The scaling exponent as a function of the system's size is shown in the right panel. The correlation length critical exponent~$\nu$ approaches a value close to the analytically predicted value $\nu=1$~\cite{ulcakar_kibble-zurek_2020}.

\subsection{Quench}
To simulate a quench we initialize the system in the trivial phase at~$u_i=-2.5$ in its ground state, assuming half-filling. The parameter~$u$ is then changed linearly in time, $u(t)=u_i+(u_f-u_i)t/\tau$, where~$u_f=-1.5$ and $\tau$ is the quench duration. The simulation ends once $t$ reaches~$\tau$.

Fig.~\ref{fig:profili}(b) shows LCM profiles during a quench. Initially, the system's evolution approximately follows the ground state~[panels~(b1) and~(b2)], before entering the freeze-out zone. After this point~[panels~(b2-5)] only the amplitude of deviations from the average changes noticeably, while the configuration of LCM inhomogeneities remains almost constant.

\begin{figure}[ht]
    \centering
    \includegraphics[width=\linewidth]{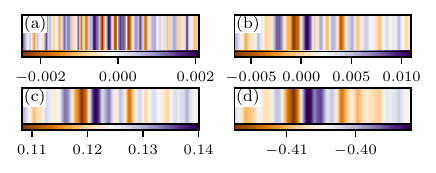}
    \caption{LCM profiles after a quench with $\tau$ equal to: 50~(a), 250~(b), 500~(c) and 1000~(d). Size of the system is $N_x=N_y=300$.}
    \label{fig:quench-profiles}
\end{figure}

Fig~\ref{fig:quench-profiles} shows LCM profiles at the end of quenches with different $\tau$. The slower the quench, the larger the LCM inhomogeneities in the final state. Note that the average LCM is not conserved exactly during the evolution for quenches with large $\tau$, despite $C$ being conserved under unitary evolution~\cite{dalessio_dynamical_2015}. This is due to finite-size effects.

\begin{figure}[ht]
    \centering
    \includegraphics[width=\linewidth]{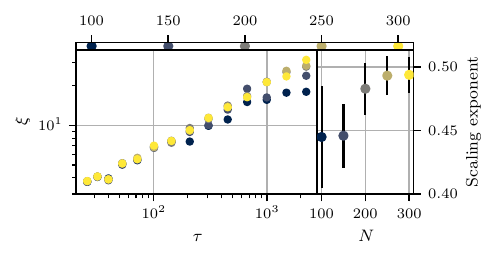}
    \caption{Left: Correlation length after slow linear quenches with different quench durations~$\tau$. Color scale represents system size~$N_x=N_y$. Right: Scaling exponent~(see Eq.~\eqref{eq:quench-xi-scaling}) obtained by fitting a power law to data shown in the left panel.}
    \label{fig:quench}
\end{figure}

The average size of inhomogeneities after a quench depends on $\tau$ as shown in Fig.~\ref{fig:quench}. Results are averaged over 100 disorder realizations. The dependence follows a power law with the scaling exponent approximately equal to $0.5$. This value coincides with the KZM prediction~\eqref{eq:quench-xi-scaling} for values of critical exponents $\nu=z=1$~\cite{ulcakar_kibble-zurek_2020}.

\section{Conclusion}\label{sec:conclusion}
In summary, we studied the local Chern and St\v{r}eda markers in Chern insulators with partial translational symmetry. We presented a numerical formula for calculating LCM in the hybrid basis both within periodic and open boundary conditions.
The Haldane model was used to investigate the behavior of LCM on a ribbon. The behavior at the boundary was found to be qualitatively different from the fully-OBC case.
We compared LCM to the local St\v{r}eda marker. The two markers were shown to give matching results, except at the boundary or when the disorder becomes very strong. 

To show the advantages of using local topological markers in the hybrid basis we used LCM to determine equilibrium and dynamical critical properties of a Chern insulator and their relation via KZM. Considering a system with partial translational symmetry allowed for the simulation of larger systems. This enabled accurate determination of scaling exponents that agree closely with analytical predictions.

Topological markers in the hybrid basis that were investigated in this paper could be used for convenient and efficient numerical calculations in many different systems that feature a translationally invariant direction. The numerical methods presented could be extended to local topological markers for other classes of topological insulators.

\acknowledgments
We acknowledge useful discussions with T.~Prijon. This work is supported by the Slovenian Research and Innovation Agency~(ARIS) under Contract No.~P1-0044 and JI-50005. J.M. also acknowledges support by ERC Consolidator Grant Boundary-101126364.

\section*{Data Availability}
The data supporting the findings of this paper are openly available~\cite{data}.

\appendix

\section{Derivation of LCM in the hybrid basis}
\label{app:derivation}
Assuming full translational symmetry, the Chern number is calculated as the integral over the 2-dimensional Brillouin zone~(BZ)
\begin{equation}\label{eq:C-BZ-def}
    C=\frac i{2\pi}\int_{\rm BZ}\dd\vb k\tr{P(\vb k)\comm{\partial_{k_x}P(\vb k)}{\partial_{k_y}P(\vb k)}},
\end{equation}
where $P(\vb k)=P(k_x,k_y)$ is the Fermi projector in the reciprocal space and the trace is over the orbital space. On a finite $N_x\times N_y$ mesh, the derivatives can be calculated using the spectral derivative
\begin{equation}
    \partial_{k_x}P(\vb k)\to\Delta_{k_x}P(\vb k)=\sum_{x=1}^{N_x}i\tilde xP(x,k_y)\frac{e^{ik_xx}}{\sqrt{N_x}},
\end{equation}
where $P(x,k_y)$ is the partial Fourier transform of $P(\vb r)$ and
\begin{equation}\label{eq:x-tilde-def}
    \tilde{x}=\begin{cases}
        \quad x\quad&\text{if}\ x<N_x/2 \\
        \quad0\quad&\text{if}\ x=N_x/2 \\
        x-N_x\quad&\text{otherwise.}
    \end{cases}
\end{equation}
After an inverse partial Fourier transform, $P(x,k_y)$ becomes $P(x,k_y)=\sum_{k_x'}P(k_x',k_y)e^{-ik_x'x}/\sqrt{N_x}$. Inserting into Eq.~\eqref{eq:C-BZ-def} and replacing the integral with a sum, we get
\begin{equation}
\begin{split}
    C=&\frac{2\pi i}{N_xN_y}\sum_{k_x,k_y}\tr\Bigg\{P(k_x,k_y) \\ &\Big[\sum_{q_x}\delta_{k_x,q_x}\sum_xi\tilde{x}\sum_{k_x'}P(k_x',k_y)\frac{e^{i(q_x-k_x')x}}{N_x}, \\ &\sum_{p_x}\delta_{q_x,p_x}\Delta_{k_y}P(k_x,k_y)\Big]\delta_{p_x,k_x}\Bigg\},
\end{split}
\end{equation}
where we have relabeled $k_x$ to $q_x$ in the first factor of the commutator and to $p_x$ in the second factor and $\delta_{k_x,q_x}=\sum_{x_1}e^{i(k_x-q_x)x_1}/N_x$ is the Kronecker delta. After expanding Kronecker deltas and some reordering of terms we get
\begin{equation}\label{eq:app-izpeljava-midstep}
    \begin{split}
        C=&\frac{2\pi i}{N_xN_y}\sum_{k_y}\tr\Bigg\{\sum_{x}\sum_{q_x}\frac{e^{iq_x(x-(x_1-x_2))}}{N_x} \\
        &\sum_{k_x}\frac{e^{ik_xx_1}}{\sqrt{N_x}}P(k_x,k_y)\frac{e^{-ik_xx_3}}{\sqrt{N_x}} \\
        &\Big[i(\widetilde{x_1-x_2}) \sum_{k_x'}\frac{e^{ik_x'x_2}}{\sqrt{N_x}}P(k_x',k_y)\frac{e^{-ik_x'x_1}}{\sqrt{N_x}}
        , \\
        &\sum_{p_x}\frac{e^{ip_xx_3}}{\sqrt{N_x}}\Delta_{k_y}P(p_x,k_y)\frac{e^{-ip_xx_2}}{\sqrt{N_x}}
        \Big]
        \Bigg\},
    \end{split}
\end{equation}
where we have made use of the identity $\sum_{q_x}e^{iq_x(x-(x_1-x_2))}/N_x=\delta_{x,x_1-x_2}$. In Eq.~\eqref{eq:app-izpeljava-midstep} we recognize the transformation of $P(k_x,k_y)$ into the hybrid basis,
\begin{equation}
    P_{x_1,x_3}(k_y)=\sum_{k_x}\frac{e^{ik_xx_1}}{\sqrt{N_x}}P(k_x,k_y)\frac{e^{-ik_xx_3}}{\sqrt{N_x}},
\end{equation}
and identify $-i(\widetilde{x_2-x_1})=-i\Delta_x$, where $\Delta_x$ is the Toeplitz matrix defined in Sec.~\ref{subsec:markers-chern}. We are left with the expression for the Chern number in the hybrid basis
\begin{equation}\label{eq:appendix-LCM}
    C=\frac{2\pi i}{N_xN_y}\sum_{k_y}\tr{P(k_y)\comm{-i\Delta_x\odot P(k_y)}{\Delta_{k_y}P(k_y)}}.
\end{equation}
The expression for LCM in this basis follows directly from Eq.~\eqref{eq:appendix-LCM}. If the difference $-i(\widetilde{x_2-x_1})$ in Eq.~\eqref{eq:app-izpeljava-midstep} is replaced with the usual difference $-i(x_2-x_1)$, then Eq.~\eqref{eq:LCM-rk} can be easily recovered.

\section{Additional profiles with disorder}\label{app:disorder}
\begin{figure}[ht]
    \centering
    \includegraphics[width=\linewidth]{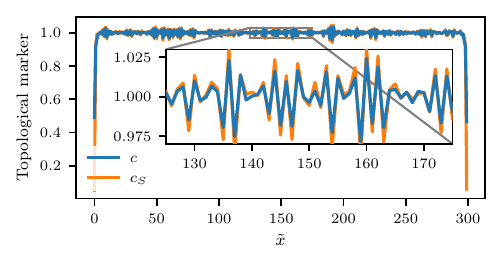}
    \caption{Configuration of LCM and local St\v{r}eda markers in a disordered Haldane ribbon. All parameters are the same as in Fig.~\ref{fig:disorder-comp}~(note the different scale), apart from $\delta\phi=\phi_0/N_y$.}
    \label{fig:disorder-comp-min-flux}
\end{figure}

The LCM and local St\v{r}eda marker profiles shown in Fig.~\ref{fig:disorder-comp}, now calculated at the magnetic flux through one unit cell $\delta\phi=\phi_0/N_y$, are shown in Fig.~\ref{fig:disorder-comp-min-flux}. The discrepancy between the markers is large at the boundaries of the system. In the bulk, the markers behave similarly.

\begin{figure}[ht]
    \centering
    \includegraphics[width=1\linewidth]{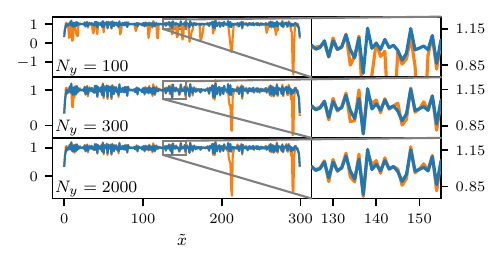}
    \caption{LCM~(blue) and local St\v{r}eda marker~(orange) for a disordered Haldane ribbon~(see Sec.~\ref{sec:haldane-ribbon}) with $\delta=3$, $\delta\phi=30\,\phi_0/N_y$ and $N_x=300$, $N_y=100,\ 300\ \text{and}\ 2000$.}
    \label{fig:disorder-app-w3}
\end{figure}

\begin{figure}[ht]
    \centering
    \includegraphics[width=1\linewidth]{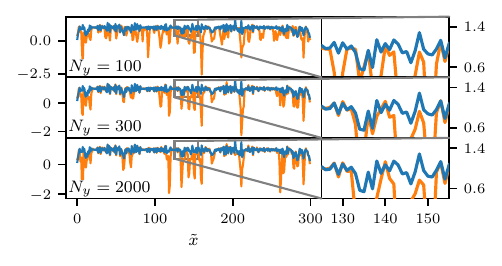}
    \caption{LCM~(blue) and local St\v{r}eda marker~(orange) for a disordered Haldane ribbon with $\delta=4$. All other parameters are exactly the same as in Fig.~\ref{fig:disorder-app-w3}.}
    \label{fig:disorder-app-w4}
\end{figure}

Figs.~\ref{fig:disorder-app-w3} and~\ref{fig:disorder-app-w4} show configurations of LCM~(blue line) and local St\v{r}eda marker~(orange line) at $\delta=3$ and 4, respectively. The left panels show systems with different sizes in the translationally invariant direction $y$: $N_y=100,\ 300\  \text{and}\ 2000$ and the right panels show a part of the system up close. For $\delta=3$~(Fig.~\ref{fig:disorder-app-w3}), the discrepancy between the markers is large everywhere when $N_y=100$. This discrepancy reduces in most points with increasing $N_y$, although the local St\v{r}eda marker still features a few large spikes, that are not present in the LCM configuration. These spikes persist at even larger $N_y$, but the agreement between the markers in other points keeps improving slightly.

For $\delta=4$~(Fig.~\ref{fig:disorder-app-w4}), there are many spikes in the local St\v{r}eda marker configuration. Consequently, the system features only a few points where the markers agree, regardless of $N_y$.

\begin{figure}[ht]
    \centering
    \includegraphics[width=0.8\linewidth]{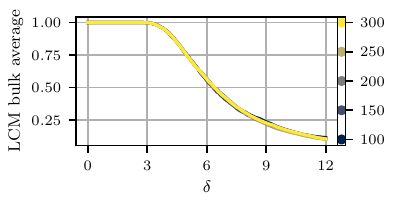}
    \caption{LCM average over the sites in the bulk at different values of $\delta$. The color scale shows system size $N_x=N_y$.}
    \label{fig:disorder-anderson}
\end{figure}

The value $\delta\gtrsim3$ corresponds to the disorder amplitude, at which the bulk average of LCM starts to deviate from the quantized $C$ of the clean system. Fig.~\ref{fig:disorder-anderson} shows the LCM bulk average~(10 sites at each edge are excluded) at different values of $\delta$ and averaged over 100 disorder realizations.

Generally, for $\delta\lesssim3$~($\delta\gtrsim3$), the behavior is qualitatively equal to the behavior at $\delta=3\ (4)$. (For $\delta<3$ local St\v{r}eda marker configurations feature fewer spikes.)
We therefore conclude that similarity between LCM and the local St\v{r}eda marker persists up to disorder amplitudes that cause a change in the system's bulk Chern number. This could be due to Anderson localization.

\bibliography{refs}

\end{document}